\documentclass[twoside,12pt]{article}

\usepackage{amssymb}
\usepackage{epsfig}
\usepackage{amsthm}
\usepackage{amsmath, mathrsfs,psfrag}

\newlength{\dinwidth}
\newlength{\dinmargin}
    
\newtheorem{theorem}{Theorem}[section]
\newtheorem{proposition}[theorem]{Proposition}
\newtheorem{definition}[theorem]{Definition}
\newtheorem{lemma}[theorem]{Lemma}

\font\fa=bbm11
\newcommand{\N}{\mbox{\fa N}}   

\newcommand{\Ibb}[1]{ {\rm I\ifmmode\mkern -3.6mu\else\kern -.2em\fi#1}}
\newcommand{\ibb}[1]{\leavevmode\hbox{\kern.3em\vrule
     height 1.2ex depth -.3ex width .2pt\kern-.3em\rm#1}}
\newcommand{\Cl}{{\ibb C}}           
\newcommand{\Rl}{{\Ibb R}}           
\newcommand{\Zl}{\mathbb{Z}}

\newcommand{\bref}[1]{(\ref{#1})}

\newcommand{\Hil}{\mathcal{H}}
 
\newcommand{\Ss}{\mathscr{S}}
\newcommand{\OO}{\mathcal{O}}
\newcommand{\Nu}{\mathscr{N}} 
\newcommand{\pol}{\mathscr{P}}

\newcommand{\te}{\theta}
\newcommand{\la}{\lambda}

\newcommand{\fti}{\widetilde{f}}

\newcommand{\zd}{z^{\dagger}}
\newcommand{\lto}{\longrightarrow}
\newcommand{\tp}[1]{^{\otimes #1}}    
\newcommand{\dom}{\mathrm{dom}}

\newcommand{\A}{\mathcal{A}}
\newcommand{\B}{\mathcal{B}}
\newcommand{\K}{\mathcal{K}}
\newcommand{\LL}{\mathcal{L}}

\title{On the Existence of Local Observables\\ in Theories With a
  Factorizing S--Matrix}
\author{Gandalf Lechner\\
Institut f\"ur Theoretische Physik, Universit\"at G\"ottingen,\\
37077 G\"ottingen, Germany\\
{\small 
{\tt lechner@physik.uni-goe.de}}}
\begin{document}
\bibliographystyle{plain}
\date{}
\maketitle

{\abstract
A recently proposed criterion for the existence of local quantum
fields with a prescribed factorizing scattering matrix is verified in
a non-trivial model, thereby establishing a new constructive
approach to quantum field theory in a particular example. The
existence proof is accomplished by analyzing nuclearity properties of
certain specific subsets of Fermionic Fock spaces.
}
\section{Introduction}

In the last few years a new strategy for the construction of
two-dimensional quantum field theories with a factorizing scattering matrix has been developed. At the
basis of this approach lies the insight of Schroer and Wiesbrock
\cite{schroer, schroer-wiesbrock} that factorizing S-matrices of
massive Bosons can be used to define Bosonic Wightman fields localized in wedge shaped
regions of two-dimensional Minkowski space by means of the
Zamolodchikov algebra \cite{zamo}. For a simple class of two-particle
scattering matrices $S_2$, it was shown in \cite{gl1} that these
semi-local fields on the one hand share many properties with a free
field, but, on the other hand, lead to non-trivial two-particle
scattering states corresponding to $S_2$. This construction opens a
new perspective in the inverse scattering problem in low-dimensional
quantum field theory, {\it i.e.} the formfactor program \cite{smirnov,
  ffp}.\\
After the construction of the wedge-local fields, a vital issue in
this approach is to show that the models so defined also contain observables localized in
{\em bounded} spacetime regions. Whereas the concrete construction of local quantum
fields turns out to be very difficult, the existence problem seems to
be more easily manageable in the algebraic setting of quantum field theory
\cite{haag}. There one considers not the wedge-local fields themselves
but rather the so-called wedge algebras generated by them. In this
framework, the existence of local observables is equivalent to the
non-triviality of certain intersections of such algebras
\cite{schroer-wiesbrock}.\\
Because of the distinguished geometric action of the modular operators
\cite{BiWi,borchers,mund} corresponding to wedge
algebras and the vacuum, these objects have been studied intensely in local quantum physics. By combining
the knowledge scattered in the literature, a convenient sufficient
condition for the existence of local
observables in wedge-local theories was recently established in
\cite{BuLe}. This criterion, known as the modular nuclearity
condition \cite{nuclearmaps1}, has previously been studied in
connection with thermodynamical properties of quantum field theories
\cite{nuclearmaps2,BiWi}.\\
Given a net $W\longmapsto
\A(W)$ of wedge algebras acting on the physical Hilbert space $\Hil$
with vacuum vector $\Omega$, consider two wedges $W_1$, $W_2$,
where $W_1$ contains the causal complement $W_2'$ of $W_2$, and the
double cone region $\OO := W_1 \cap W_2$. The content of the modular
nuclearity condition is the following: If the map
\begin{equation}\label{mnc}
  \Xi:\A(W_2')\lto\Hil,\qquad \Xi(A) := \Delta_{W_1}^{1/4}A\Omega,
\end{equation}
is nuclear\footnote{See Definition 2.1}, non-trivial operators
localized in $\OO$ do exist \cite{BuLe}. Here $\Delta_{W_1}$ denotes the modular operator of
$(\A(W_1),\Omega)$, which in the models considered acts
simply as a boost with an imaginary rapidity parameter \cite{BuLe}.\\
Although this criterion does not solve the task of the explicit
construction of local operators, it opens up the possibility to decide
whether such fields exist. Moreover, it provides information about the structure of local algebras determined by them. We are therefore led to the question whether the maps
\bref{mnc} are nuclear, for example in the class of
S-matrices considered in \cite{gl1}.\\
As a first step in this direction, we verify in the present Letter the
modular nuclearity condition in an explicit example of a factorizing
theory. The model chosen is fixed by the constant two-particle scattering matrix
$S_2=-1$. It is thus related to the Ising model in the scaling limit, above
the critical temperature \cite{BKW}. The underlying fields are most conveniently
represented on an antisymmetric Fock space. Because of this formal analogy
to systems of free Fermions, it is possible to study nuclearity
properties of maps like \bref{mnc} in  a mathematical
framework wide enough to cover interaction-free Fermionic theories as
well. This has the advantage that, as a byproduct of our present investigation, we can also show that the energy
nuclearity condition of Buchholz and Wichmann \cite{BuWi} is satisfied
in theories describing free Fermions. Although this was expected from
the thermodynamical interpretation of the energy nuclearity condition,
only Bosonic theories were shown to satisfy this criterion up to now
\cite{BuWi, BuJa}.\\
This article is organized as follows. 
The analysis of nuclearity properties of maps on Fermionic Fock space
in a general setting is presented in section two. In section
three, we verify the modular nuclearity condition in the
factorizing theory based on the two-particle S-matrix $S_2=-1$ by
applying these results. Some comments about the energy
nuclearity condition for free Fermions are given in the Conclusions.

\section{Nuclear Maps on Fermionic Fock Space}
\setcounter{equation}{0}
In this section we study nuclearity properties of certain subsets of
antisymmetric Fock space in a general setting. The results obtained
here will subsequently be applied to quantum field theoretic models.\\
The mathematical structure needed for our analysis is the following:
Let $\K$ be a complex Hilbert space with an antilinear involution
$\Gamma = \Gamma^* = \Gamma^{-1}$ acting on it. (In the applications, $\K$ will be realized as a one particle space of square
integrable functions on the upper
mass shell, and $\Gamma$ corresponds to complex conjugation in
configuration space.) We consider two closed, complex
subspaces $\LL_\varphi$ and $\LL_\pi$ of $\K$ which are invariant
under $\Gamma$ and the real linear subspace defined by
\begin{equation}
  \label{L}
  \LL := (1+\Gamma)\LL_\varphi + (1-\Gamma)\LL_\pi \;.
\end{equation}
By second quantization one obtains the antisymmetric Fock space $\Hil$
over $\K$, the vacuum vector $\Omega\in\Hil$ and the usual
annihilation and creation operators $a(\psi)$ and $a^*(\psi) =
a(\psi)^*$, $\psi\in\K$, representing the CAR algebra on $\Hil$, {\em
  i.e.} ($\psi_1,\psi_2\in\K$)
\begin{eqnarray}
  [a(\psi_1),a(\psi_2)]_+ &=& 0,\label{CAR1}\\
  \mbox{[}a(\psi_1),a^*(\psi_2)]_+ &=& \langle \psi_1,\psi_2 \rangle \cdot 1\,.\label{CAR2}
\end{eqnarray}
Here we introduced the notation $[A,B]_\pm =
AB \pm BA$ for the (anti-) commutator and $\langle\,.\,,\,.\,\rangle$
for the scalar product on $\K$. (The scalar product on $\Hil$ will be
denoted by the same symbol.) We adopt the convention that the
creation operator $a^*(\psi)$ depends complex linearly on
$\psi\in\K$. The CAR
relations imply that the annihilation and creation operators are
bounded \cite{BraRob2}: $\|a(\psi)\|=\|a^*(\psi)\|=\|\psi\|$.\\
Furthermore, we introduce a fermionic field operator 
\begin{equation}\label{field}
  \phi(\psi) := a^*(\psi) + a(\psi)\,,\qquad\quad\psi\in\LL\,,
\end{equation}
as well as the auxiliary fields ($\psi\in\K$)
\begin{eqnarray}
  \varphi(\psi) := a^*(\psi)+a(\Gamma\psi)\,,\qquad\quad
  \pi(\psi) := i(a^*(\psi)-a(\Gamma\psi))\,,
\end{eqnarray}
which are related to the time zero Cauchy data of $\phi$ in the field
theoretic context. Note that $\varphi(\psi)^*=\varphi(\Gamma\psi)$,
$\pi(\psi)^*=\pi(\Gamma\psi)$ and that
$\varphi(\psi_1)$ and $\pi(\psi_2)$ anticommute for arbitrary
$\psi_1,\psi_2\in\K$. For later use we also state 
\begin{equation}
  \label{a}
  a(\Gamma\psi) = \frac{1}{2}\left(\varphi(\psi)+i\pi(\psi)\right)\,.
\end{equation}
The field $\phi$ generates the von Neumann algebra 
\begin{eqnarray}
  \A(\LL) := \left\{\phi(\psi)\,:\,\psi\in\LL\right\}''\,,
\end{eqnarray}
and we assume that the vacuum vector $\Omega$ is seperating for this
algebra\footnote{As in \cite{BuLe,gl1}, one may equivalently write
  $\A(\LL)=\{\exp(i\phi(\psi))\,:\,\psi\in\LL\}''$ \cite{kadring1}.}.

The last element needed for our analysis is a densely defined,
strictly positive operator $X$ on $\K$, which commutes with the
involution $\Gamma$. In particular, $X$ is assumed
to be invertible. Having in mind the nuclearity conditions mentioned in the
Introduction, one should think of $X$ as representing one of the
following two operators: In connection with the modular nuclearity
condition, put $X=\Delta^{1/4}$, where $\Delta$ is the
modular operator of some von Neumann algebra containing $\A(\LL)$
with respect to the vacuum vector, and in the context of the energy
nuclearity condition, put $X=e^{-\beta H}$, where $\beta>0$ is the inverse
temperature and $H$ the Hamiltonian of the theory. As the former
example indicates, $X$ is not required to be bounded. 
We use the same symbol $X$ to denote its second quantization $\bigoplus_{n=0}^\infty X\tp{n}$ and assume that $\A(\LL)\Omega$ is
contained in its domain.\\
It is our aim to find sufficient conditions on the real subspace $\LL$
and the operator $X$ that imply the nuclearity of the map
\begin{eqnarray}\label{nuc-map}
  \Xi_\LL : \A(\LL) \lto \Hil\,,\qquad\qquad\Xi_\LL(A) := XA\Omega\,.
\end{eqnarray}
For the convenience of the reader, we briefly recall the notion of a
nuclear map between two Banach spaces ({\em cf}., for example, \cite{pietsch}).
\begin{definition}
A linear map $\Xi$ between two Banach spaces $\A$ and $\Hil$ is said to be
nuclear if there exists a sequence of linear functionals $\rho_k\in\A^*$,
$k\in\N$, and a sequence of vectors $\Psi_k\in\Hil$, $k\in\N$, such that for all $A\in\A$
\begin{eqnarray}\label{nuc-rep}
  \Xi(A) = \sum_{k=1}^\infty \rho_k(A)\cdot\Psi_k,\qquad
  \sum_{k=1}^\infty \|\rho_k\|_{\A^*}\|\Psi_k\|_\Hil
  < \infty
\end{eqnarray}
The nuclear norm $\|\Xi\|_1$ of such a map is defined as
\begin{eqnarray}\label{nuc-norm}
  \|\Xi\|_1 := \inf_{\rho,\phi}\sum_{k=1}^\infty \|\rho_k\|_{\A^*}\|\Psi_k\|_\Hil\,,
\end{eqnarray}
where the infimum is taken with respect to all sequences
$\rho_k\in\A^*$, $\Psi_k\in\Hil$, $k\in\N$, complying with the above
conditions.\\
\end{definition}

As $\Omega$ seperates $\A(\LL)$ and $X$ is invertible, the nuclearity
of the map $\Xi_\LL$ is equivalent to the nuclearity of the set
\begin{eqnarray}\label{nuc-set}
  \Nu(X,\LL) := \left\{XA\Omega\,:\, A\in\A(\LL),\,\|A\|\leq 1\right\}^-\,,
\end{eqnarray}
which is a subset of $\Hil$ (the bar indicates closure in the norm topology
of $\Hil$), and the nuclearity index of this set \cite{pietsch} coincides
with the nuclear norm of $\Xi_\LL$. We may thus treat the map
\bref{nuc-map} and the set \bref{nuc-set} on an equal footing.\\
Denoting by $E_{\varphi}, E_\pi \in \B(\K)$ the orthogonal projections onto
$\LL_{\varphi}$, $\LL_\pi$, respectively, the nuclearity properties of
\bref{nuc-map} are characterized in the following Proposition.
\begin{proposition}\label{prop}
  Assume that $E_\varphi X$ and $E_\pi X$ extend to 
  trace class operators on $\K$. Then $\Xi_\LL$ is a
  nuclear map, and its nuclear norm is bounded by
  \begin{equation}\label{xi-bound}
  \|\Xi_\LL\|_1 \leq
  e^{2\|E_\varphi X\|_1}\cdot e^{2\|E_\pi X\|_1}\,.
  \end{equation}
\end{proposition}
In comparison with the analogous result for Bosons \cite[Theorem
2.1]{BuJa} one notices two differences: Firstly, the conditions on
$E_\varphi X$, $E_\pi X$ are relaxed since the bounds $\|E_\varphi
X\|<1$, $\|E_\pi X\|<1$ on their
operator norms are not required here. Secondly, our bound on the
nuclearity index is smaller than the corresponding one for Bosons,
$\det(1-|E_\varphi X|)^{-2}\cdot \det(1-|E_\pi X|)^{-2}$, obtained in \cite{BuJa}. This can be
seen from the following simple inequality, valid for any non-zero trace class
operator $T$ with norm $\|T\| < 1$. The singular values of
$T$ are denoted by $t_n$, repeated according to multiplicity.
\begin{equation*}
  e^{2\|T\|_1} = e^{2\sum_{n=1}^\infty |t_n|}
  =
    \prod_{n=1}^\infty \left(e^{-|t_n|}\right)^{-2}
    < \prod_{n=1}^\infty (1-|t_n|)^{-2}
    = \det(1-|T|)^{-2}\,.
\end{equation*}
This result is due to the Pauli principle; it may be understood in
analogy to the difference between the partition functions of the
non-interacting Bose and Fermi gases in the grand canonical ensemble.\\

The rest of this section is devoted to the proof of Proposition
\ref{prop}. In a first step, we proceed to the
polynomial algebra generated by the field,
\begin{equation}\label{pol}
  \pol(\LL) := \mbox{span}\{\phi(\psi_1)\cdots\phi(\psi_n)\,:\, n\in\N\,,\psi_i\in\LL\}\,.
\end{equation}
As $\phi(\psi)$ is bounded, $\pol(\LL)$ is a weakly dense subalgebra
of $\A(\LL)$. In view of the closedness of $X$, we may apply
Kaplansky's density theorem and conclude that if the set
\begin{eqnarray}\label{defn0}
  \Nu_0(X,\LL) :=   \left\{XA\Omega\,:\,A\in\pol(\LL),\|A\|\leq 1 \right\}^-
\end{eqnarray}
is nuclear, then the larger set \bref{nuc-set} is nuclear, too, with the
same nuclearity index \cite{BuJa}. It is therefore sufficient to study
the restriction of $\Xi_\LL$ to $\pol(\LL)$, and we begin with some
comments about this algebra.\\
The polynomial algebra has the structure of
a $\Zl_2$-graded $*$-algebra, with the even and odd parts $\pol^+(\LL)$
and $\pol^-(\LL)$ given by the linear span of the field monomials of
even and odd order, respectively. On $\pol(\LL)$ acts the
grading automorphism
\begin{equation}
  \gamma(A^++A^-) := A^+-A^-\,,\qquad A^\pm \in \pol(\LL)^\pm\,.
\end{equation}
As $\|\gamma\|=1$ and $A^\pm = \frac{1}{2}(A\pm\gamma(A))$, we
conclude $\|A^\pm\|\leq\|A\|$.\\

The following Lemma about the interplay of the CAR algebra and
$\pol(\LL)$ in connection with the real linear structure of $\LL$
is the main technical tool in the proof of Proposition \ref{prop}. We
will denote the symplectic complement of $\LL$ by 
\begin{equation}
\LL'=\{\psi\in\K\,:\,\langle\psi,\xi\rangle=\langle\xi,\psi\rangle\quad\forall\xi\in\LL\}. 
\end{equation}
In preparation recall that an odd
derivation on a $\Zl_2$-graded algebra $\pol$ is a linear map
$\delta:\pol\to\pol$ which satisfies $\delta(\pol^\pm) \subset \pol^\mp$
and obeys the graded Leibniz rule
\begin{equation}\label{leibniz}
  \delta(A^\pm B) = \delta(A^\pm)B \pm A^\pm\delta(B),\qquad\quad
  A^\pm\in\pol^\pm\,,\quad B\in\pol.
\end{equation}
\begin{lemma}\label{Lemma1}
  For arbitrary $\psi\in\K$, the assignments
  \begin{eqnarray}\label{def-delta}
    \delta^\pm_\psi(A) &:=&
    \tfrac{1}{2}\left[\varphi((1\mp\Gamma)\psi)+i\pi((1\pm\Gamma)\psi),A^+\right]_-\\
     && + \tfrac{1}{2}\left[\varphi((1\mp\Gamma)\psi)+i\pi((1\pm\Gamma)\psi),A^-\right]_+\nonumber
  \end{eqnarray}
  define two odd derivations on $\pol(\LL)$ which are real linear in $\psi$. These maps satisfy the bounds
  \begin{eqnarray}
    \|\delta^+_\psi(A^\pm)\| &\leq& \left(\|(1 -
      \Gamma)E_\varphi\psi\|^2 + \|(1 +
      \Gamma)E_\pi\psi\|^2\right)^{1/2}\cdot\|A^\pm\|,\label{d-norm}\\
     \|\delta^-_\psi(A^\pm)\| &\leq& \left(\|(1 +
      \Gamma)E_\varphi\psi\|^2 + \|(1 - \Gamma)E_\pi\psi\|^2\right)^{1/2}\cdot\|A^\pm\|\,.\label{d-norm2}
  \end{eqnarray}
  Moreover, if $\psi\in\LL'$,
  \begin{eqnarray}\label{kill}
    \delta^+_\psi=0\,,\qquad\qquad \delta^-_{i\psi}=0\,.
  \end{eqnarray}
\end{lemma}
\begin{proof}
  The real linearity of $\psi\longmapsto\delta^\pm_\psi$ follows
  directly from the definition \bref{def-delta} and the real linearity of
  $\varphi,\pi$ and $\Gamma$.\\
  As $\delta^\pm_\psi$ are complex linear maps on
  $\pol(\LL)$, it suffices to consider their action on field monomials 
  $\phi(\xi_1)\cdots\phi(\xi_n)$, $\xi_1,...,\xi_n\in \LL$ to prove
  the other assertions of the Lemma. We also write
  $\phi_k := \phi(\xi_k)$ and carry out a proof based on induction in the field
  number $n$. For $n=1$, the CAR relations (\ref{CAR1},\ref{CAR2}) imply that
  \begin{eqnarray}
    \delta^\pm_\psi(\phi(\xi)) &=&
    \tfrac{1}{2} \left[
      \varphi((1\mp\Gamma)\psi)
      + i\pi((1\pm\Gamma)\psi),\phi(\xi)\right]_+\nonumber\\
    &=& \tfrac{1}{2}
    \left(\langle \xi,(1\mp\Gamma)\psi\rangle
    \mp \langle (1\mp\Gamma)\psi,\xi\rangle
    -\langle\xi,(1\pm\Gamma)\psi\rangle 
    \pm \langle(1\pm\Gamma)\psi,\xi\rangle
  \right)\cdot 1\nonumber\\
  &=& \left( \langle\Gamma\psi,\xi\rangle \mp
    \langle\xi,\Gamma\psi\rangle\right)\cdot 1.\label{calc1}
  \end{eqnarray}
  As $\LL$ is $\Gamma$-invariant, so is $\LL'$, and hence
  $\psi\in\LL'$ implies $\delta^+_\psi(\phi(\xi))=0$,
  $\delta^-_{i\psi}(\phi(\xi))=0$. Being a multiple of the identity,
  $\delta^\pm_\psi(\phi(\xi))$ is contained in $\pol^+(\LL)$ for arbitrary $\psi\in\K$. The step from $n$ to $n+1$ fields is achieved by considering
  \begin{eqnarray}\label{recurse}
    \begin{array}{rcl}
    [F,\phi_1\cdots\phi_{2n}]_-
    &=& 
    [F,\phi_1\cdots\phi_{2n-1}]_+\cdot\phi_{2n}
    -
    \phi_1\cdots\phi_{2n-1}\cdot[F,\phi_{2n}]_+\,,\\\vspace*{-3mm}&&\\
    \mbox{[}F,\phi_1\cdots\phi_{2n+1}]_+
    &=& 
    [F,\phi_1\cdots\phi_{2n}]_-\cdot\phi_{2n+1}
    +
    \phi_1\cdots\phi_{2n}\cdot[F,\phi_{2n+1}]_+\,,
    \end{array}
  \end{eqnarray}
  with
  $F=\frac{1}{2}(\varphi((1\mp\Gamma)\psi)+i\pi((1\pm\Gamma)\psi))$. It follows from these formulae inductively that $\delta^\pm_\psi$ turn even
  elements of $\pol(\LL)$ into odd ones and vice versa. Moreover,
  $\delta^+_\psi=0$, $\delta^-_{i\psi}=0$ for $\psi\in\LL'$ because of
  the corresponding result for $n=1$. By direct
  calculation, one can also verify the Leibniz rule \bref{leibniz}.
  We have
  thus shown that $\delta^\pm_\psi$ are odd derivations of
  $\pol(\LL)$ satisfying \bref{kill}.\\
  To prove the norm estimate \bref{d-norm}, we first note that
  \begin{equation}
    \psi':=\left(\tfrac{1}{2}(1+\Gamma)(1-E_\pi)+\tfrac{1}{2}(1-\Gamma)(1-E_\varphi)\right)\psi\,.
  \end{equation}
  is an element of the symplectic complement $\LL'$ for arbitrary $\psi\in\K$,
  as can be easily verified using \bref{L}. Since $\delta^+_{\psi'}=0$ and $\delta^+_\psi$ is real linear in $\psi$, we have
  \begin{eqnarray}
    \|\delta^+_\psi(A^\pm)\| &=& \|\delta^+_{\psi-\psi'}(A^\pm)\|\nonumber\\
    &=& 
    \tfrac{1}{2} \|
    [\varphi((1-\Gamma)E_\varphi\psi)+i\,\pi((1+\Gamma)E_\pi\psi),A^\pm]_\mp\|\nonumber\\
    &\leq&\|\varphi((1-\Gamma)E_\varphi\psi)+i\,\pi((1+\Gamma)E_\pi\psi)\|\cdot\|A^\pm\|\,.\label{d-norm-1}
  \end{eqnarray}
  To proceed to the estimate \bref{d-norm}, let
  $\chi_-:=(1-\Gamma)E_\varphi\psi$, $\chi_+ := (1+\Gamma)E_\pi\psi$. As $(\varphi(\chi_-)+i\pi(\chi_+))^* = -(\varphi(\chi_-)+i\pi(\chi_+))$ and $\varphi(\chi_-)$ anticommutes with $\pi(\chi_+)$,
  \begin{eqnarray*}
    \|\varphi(\chi_-)+i\pi(\chi_+)\| = \|\varphi(\chi_-)^2 -
    \pi(\chi_+)^2\|^{1/2}
    = \left(\|\chi_-\|^2 + \|\chi_+\|^2\right)^{1/2}.
  \end{eqnarray*}
  Together with \bref{d-norm-1} this implies
  the claimed norm bound \bref{d-norm} for $\delta^+_\psi$. To
  establish the corresponding inequality \bref{d-norm2} for
  $\delta^-_\psi$, consider the vector
  \begin{eqnarray}
    \psi'':=\left(\tfrac{1}{2}(1-\Gamma)(1-E_\pi)+\tfrac{1}{2}(1+\Gamma)(1-E_\varphi)\right)\psi\,,
  \end{eqnarray}
  which is contained in $i\LL'$ for any $\psi\in\K$.
  The norms of $\delta^-_\psi(A^\pm) = \delta^-_{\psi-\psi''}(A^\pm)$
  can then be estimated along the same lines as before.
\end{proof}

After these preparations, we now turn to the proof of the nuclearity of $\Xi_\LL$ by estimating
the size of its image in $\Hil$. Let $\xi_1,...,\xi_n\in\K \cap \dom(X)$ and
$A\in\pol(\LL)$. In view of the second quantization structure
of $X$ and the annihilation property of $a(\xi_j)$, we have
\begin{eqnarray}\label{f1}
\lefteqn{\langle a^*(\Gamma\xi_1)\cdots a^*(\Gamma\xi_n)\Omega,XA^\pm\Omega\rangle =
  \langle \Omega,\, a(X\Gamma\xi_n)\cdots a(X\Gamma\xi_1)A^\pm\Omega\rangle}\\
  && \qquad\qquad\;\qquad\qquad= \langle \Omega,\,[a(X\Gamma\xi_n),[\,...\,[a(X\Gamma\xi_2),[a(X\Gamma\xi_1),A^\pm]_\mp]_\pm]\,...\,]_\pm]_\mp\Omega\rangle\,.\nonumber
\end{eqnarray}
From the inside to the outside, commutators and anticommutators are
applied alternatingly. We start with a commutator
$[a(X\Gamma\xi_1),A^+]_-$ if $A=A^+$ is even and with an anticommutator
$[a(X\Gamma\xi_1),A^-]_+$ if $A=A^-$ is odd. Writing the annihilation
operator as a linear combination of the auxiliary fields \bref{a} and
recalling that $X$ commutes with $\Gamma$, one notes that
the innermost (anti-) commutator is
\begin{equation}
[a(X\Gamma\xi_1),A^\pm]_\mp = \frac{1}{2}(\delta_{X\xi_1}^+ + \delta_{X\xi_1}^-)(A^\pm)\,.
\end{equation}
Making use of this equality for all of the $n$ (anti-) commutators, it becomes apparent
that \bref{f1} can be rewritten as
\begin{equation}\label{d-eqn}
\langle a^*(\Gamma\xi_1)\cdots
a^*(\Gamma\xi_n)\Omega,XA^\pm\Omega\rangle =
2^{-n}\,\langle\Omega, ((\delta^+_{X\xi_n}+\delta^-_{X\xi_n})\cdots(\delta^+_{X\xi_1}+\delta^-_{X\xi_1})(A^\pm))\Omega\rangle\,.
\end{equation}
According to the assumptions of Proposition \ref{prop}, 
\begin{eqnarray}
  T_\varphi := E_\varphi X\,, \qquad\qquad T_\pi := E_\pi X\,,
\end{eqnarray}
are trace class operators on $\K$. Taking into account that
$\delta^\pm_{X\xi_j}$ are odd derivations on $\pol(\LL)$, an
application of the bounds (\ref{d-norm}, \ref{d-norm2}) to \bref{d-eqn} yields
\begin{eqnarray*}
  |\langle a^*(\Gamma\xi_1)\cdots
  a^*(\Gamma\xi_n)\Omega,XA^\pm\Omega\rangle|
  &\leq& 
  2^{-n}\prod_{j=1}^n 
  \big(\!
    \left(
      \|(1 - \Gamma)T_\varphi\xi_j\|^2 + \|(1 +\Gamma)T_\pi \xi_j\|^2
    \right)^\frac{1}{2}\\
    &&\!\!\!\!\!+
    \left(
      \|(1+ \Gamma)T_\varphi \xi_j\|^2 + \|(1 -\Gamma)T_\pi\xi_j\|^2
    \right)^\frac{1}{2}
  \big)\cdot\|A^\pm\|\,.
\end{eqnarray*}
Following \cite{BuWi,BuJa} we now consider the positive operator
\begin{eqnarray}
  T := \left(|T_\varphi|^2 + |T_\pi|^2 \right)^{1/2}
\end{eqnarray}
which is in the trace class, too, satisfies $\|T\|_1 \leq
\|T_\varphi\|_1 + \|T_\pi\|_1$ \cite{kosaki} and commutes with
$\Gamma$ since $T_\varphi$ and $T_\pi$ do. As $T^2 \geq
|T_\varphi|^2$, $T^2\geq |T_\pi|^2$,
\begin{equation*}
  \|\tfrac{1}{2}(1 \mp \Gamma)T_\varphi\xi_j\|^2 + \|\tfrac{1}{2}(1
  \pm\Gamma)T_\pi \xi_j\|^2 \leq  \|\tfrac{1}{2}(1 \mp
  \Gamma)T\xi_j\|^2 + \|\tfrac{1}{2}(1  \pm\Gamma)T \xi_j\|^2 = \|T\xi_j\|^2\,.
\end{equation*}
In terms of $T$, we thus arrive at the estimate
\begin{eqnarray}\label{est-T}
  |\langle a^*(\Gamma\xi_1)\cdots a^*(\Gamma\xi_n)\Omega, XA^\pm\Omega\rangle|
  \leq 2^n\,\|A^\pm\| \cdot\prod_{j=1}^n \|T\xi_j\|\,.
\end{eqnarray}
Although this bound was derived for $\xi_1,...,\xi_n\in\K\cap\dom(X)$
only, it holds for arbitrary $\xi_1,...,\xi_n\in\K$ since
$\K\cap\dom(X)\subset\K$ is dense and the left- and right hand sides
of \bref{est-T} are continuous in the $\xi_j$. With the estimate
\bref{est-T}, we are now able to give a bound on the nuclearity index of
the set \bref{nuc-set}.
 
The positive trace class operator $T$ acts on $\psi\in\K$ as $T\psi =
\sum_{k=1}^\infty t_k\langle b_k,\psi\rangle b_k$, where $b_k\,,k\in\N$, is an orthonormal basis of
$\K$ and $t_k$ the (positive) eigenvalues of $T$, repeated according
to multiplicity, {\em i.e.} $\sum_{k=1}^\infty t_k = \|T\|_1
< \infty$. Moreover, since $\Gamma$ and $T$ commute, we may choose the
basis vectors $b_k$ to be eigenvectors of $\Gamma$ as well. As a consequence of the Pauli principle, the vectors
\begin{equation}\label{basis-n}
  b_{\bf k} := a^*(\Gamma b_{k_1})\cdots a^*(\Gamma b_{k_n})\Omega = \pm\, a^*(b_{k_1})\cdots a^*(b_{k_n})\Omega\,,
\end{equation}
form an orthonormal basis of the totally antisymmetric subspace of
$\K\tp{n}$ (the fermi\-onic $n$-particle space) if the multi-index ${\bf
k}:=(k_1,...,k_n)$ varies over $k_1<k_2<...<k_n$, $k_1,...,k_n\in\N$.\\
Note
that $XA\Omega$ has even (odd) particle number if $A\in\pol(\LL)$ is
even (odd). By the Fock structure of $\Hil$, we have for each
$\Xi_\LL(A) = XA\Omega \in \Nu(X,\LL)$ the decomposition  
\begin{eqnarray}
  \Xi_\LL(A) = \sum_{n=0}^\infty \sum_{k_1<..<k_{2n}} \langle
    b_{\bf k},XA^+\Omega\rangle\cdot b_{\bf k} + \sum_{n=0}^\infty
    \sum_{k_1<..<k_{2n+1}} \langle
    b_{\bf k},XA^-\Omega\rangle\cdot b_{\bf k}\,,
\end{eqnarray}
as an example for a representation of the type \bref{nuc-rep} of
$\Xi_\LL$. As $\|b_{\bf k}\|=1$ for all $k_1,...,k_n\in\N$, and $\|A^\pm\|\leq\|A\|$, the
sum of the expansion coefficients can be estimated with the help of
\bref{est-T} as follows:
\begin{eqnarray}
\lefteqn{\sum_{n=0}^\infty \bigg(
  \sum_{1\leq k_1<...<k_{2n}} |\langle b_{\bf k},XA^+\Omega\rangle| 
  +
  \sum_{1\leq k_1<...<k_{2n+1}} |\langle b_{\bf
    k},\,XA^-\Omega\rangle|\bigg)\nonumber}\\
&\leq& \|A^+\|\sum_{n=0}^\infty 2^{2n}\!\!
  \sum_{1\leq k_1<...<k_{2n}}\prod_{j=1}^{2n} \|T b_{k_j}\|
  +
  \|A^-\|\sum_{n=0}^\infty 2^{2n+1}\!\! \sum_{1\leq k_1<...<k_{2n+1}} \prod_{j=1}^{2n+1}\|T b_{k_j}\|\nonumber\\
&\leq& \|A\|\cdot \sum_{n=0}^\infty\, \sum_{1\leq k_1<...<k_n}
\prod_{j=1}^n \,2 t_{k_j}\label{part}\;.
\end{eqnarray}
According to \bref{nuc-norm}, the sum \bref{part} provides an upper
bound for the nuclear norm of $\Xi_\LL$. To compute this sum, note
that \bref{part} is nothing else but the partition function of the ideal Fermi
gas with Hamiltonian $e^{-\beta H}=2T$ and zero chemical potential
in the grand canonical ensemble. This leads to the estimate ({\em cf}., for example, \cite{BraRob2})
\begin{eqnarray}
  \|\Xi_\LL\|_1 \leq \sum_{n=0}^\infty\, \sum_{1\leq k_1<...<k_n}
\prod_{j=1}^n \,2 t_{k_j} = \prod_{j=1}^\infty (1+2t_j) = \det(1+2T)\,.
\end{eqnarray}
As $\det(1+2T)\leq \exp(2\|T\|_1) < \infty$, the nuclearity of $\Xi_\LL$ follows. Taking into account 
\begin{eqnarray}
  \|T\|_1 \leq \|T_\varphi\|_1 + \|T_\pi\|_1 = \|E_\varphi X\|_1 +
  \|E_\pi X\|_1,
\end{eqnarray}
we also obtain the bound \bref{xi-bound} given in the Proposition. {\hfill $\square$ \\[2mm] \indent}

\section{Application to Factorizing Theories}
\setcounter{equation}{0}

Our main interest in Proposition \ref{prop} is based on its
importance for the modular nuclearity condition in a model of Bose
type particles of mass $m$ whose interaction is described by the S-matrix 
\begin{equation}\label{S}
  S = (-1)^{N(N-1)/2},
\end{equation}
where $N$ denotes the particle number operator. This model is related to the scaling limit of the
two-dimensional Ising model above the critical temperature
\cite{sato}. The form factors of the field appearing in this approach
have been computed in \cite{BKW}, and the $n$-point functions in
\cite{mccoy,sato,schroer-truong}. Although this model has been thoroughly discussed, an investigation from the algebraic point of view taken
here is interesting in two respects: Firstly, no rigorous proof of the
Wightman axioms for the computed $n$-point functions is known to
us. Secondly, this model is the first example of a quantum field
theory with non-trivial S-matrix constructed by the approach described in the Introduction. As
the modular nuclearity condition is only a sufficient criterion for
the existence of local operators, it is an important test if it is
satisfied here. By Proposition \ref{prop}, the verification of this
condition can be simplified to a problem on the one particle space.\\
We begin by briefly recalling the structure of the models under
consideration in a manner adapted to the discussion in section
two. For a more thorough treatment, see \cite{gl1, BuLe}.\\
The net structure of
the algebras of observables localized in wedges $W$ arises from a net
of symplectic subspaces of the one particle space by a ``second quantization''
procedure. Recall that in two dimensions, all wedges are translates of the right wedge
\begin{equation}
    W_R := \{x\in\Rl^2\,:\,x_1>|x_0|\}
\end{equation}
or of its causal complement $W_L=-W_R$.\\
The one particle Hilbert space $\K$ can be identified with
$L^2(\Rl,d\te)$ by using the rapidity parametrization of the upper
mass shell, $p(\te) := m(\cosh\te,\,\sinh\te)$. On $\K$ we have a
unitary positive energy represenation $U$ of the proper orthochronous Poincar\'e
group defined as follows: Let $B(\la)$ denote a proper Lorentz transformation with rapidity $\la$
and $x\in\Rl^2$ a translation. We set
\begin{eqnarray}
  (U(x,B(\la))\psi)(\te) := e^{ip(\te)x}\cdot\psi(\te-\la)\,.
\end{eqnarray}
We furthermore introduce the notation $\omega$ for the one-particle
Hamiltonian which acts on a dense domain in $\K$ by
multiplication with $m\cosh\te$. With the help of the involution
\begin{eqnarray}
  (\Gamma\psi)(\te) := \overline{\psi(-\te)}
\end{eqnarray}
and the auxiliary, $\Gamma$-invariant spaces
\begin{eqnarray}
  \LL_\pm :=   \{\te\mapsto\fti(m\sinh\te)\,:\,f\in\Ss(\Rl_\pm)\}\,,
\end{eqnarray}
one defines
\begin{eqnarray}\label{lx}
  \LL_\varphi(W_L+x) &:=& \{U(x)\LL_-\}^-,\qquad\quad
  \LL_\pi(W_L+x) := \{U(x)\omega\,\LL_-\}^-\,,\\
  \LL_\varphi(W_R+x) &:=& \{U(x)\LL_+\}^-,\qquad\quad
  \LL_\pi(W_R+x) :=  \{U(x)\omega\,\LL_+\}^-\,,
\end{eqnarray}
and the real linear subspaces $\LL(W)\subset\K$ in accordance with the
procedure in section two as 
\begin{eqnarray}
\LL(W) &:=& (1+\Gamma)\LL_\varphi(W) + (1-\Gamma)\LL_\pi(W)\,.  
\end{eqnarray}
The full Hilbert space $\Hil$ of this model is the Fermionic Fock
space over $\K$. This is a special case in the class of factorizing S-matrices
considered in \cite{gl1}, where one has a represention of Zamolodchikov's algebra
\begin{eqnarray}
  \zd(\te_1)\zd(\te_2) &=& S_2(\te_1-\te_2)\zd(\te_2)\zd(\te_1),\\
  z(\te_1)\zd(\te_2) &=& S_2(\te_2-\te_1)\zd(\te_2)z(\te_1) +
  \delta(\te_1-\te_2)\cdot 1,
\end{eqnarray}
by operator valued distributions $z(\te),\zd(\te)$ acting on a Fock
space with a $S_2$-dependent symmetry
structure \cite{liguori}. The so-called scattering function $S_2$
appearing here is closely related to the two-particle S-matrix.\\
As the S-matrix \bref{S} corresponds to the scattering function
$S_2(\te)=-1$, the Hilbert space $\Hil$ is in this case the Fermionic Fock space
over $\K$ and one may define annihilation and creation operators
representing the CAR algebra (\ref{CAR1}, \ref{CAR2}) as
\begin{equation*}
  a^*(\psi) := \zd(\psi) = \int d\te\, \psi(\te)\zd(\te),\quad
  a(\psi) := z(\overline{\psi}) = \int
  d\te\, \overline{\psi(\te)}z(\te)\,\qquad \psi\in \K\,.
\end{equation*}
Furthermore, the wedge-local field operator considered in \cite{gl1}
has the same form as the field $\phi(\psi)$ defined in \bref{field}.
Note that we do {\em not\/} deal here with a free
Fermionic field, but rather with an interacting Bose field represented
on an auxiliary antisymmetric Fock space. As a matter of fact, all factorizing
models considered in \cite{gl1} have Bosonic scattering states.\\
By means of $\phi$, one can construct a wedge-dual net of von Neumann algebras from the subspaces $\LL(W)$ as
\begin{eqnarray}\label{aw}
  \A(W_L+x) := \{\phi(\psi)\,:\,\psi\in\LL(W_L+x)\}'',\qquad
  \A(W_R+x) := \A(W_L+x)'.
\end{eqnarray}
This net is covariant with respect to (the second quantization of)
$U$, and the Fock vacuum vector
$\Omega\in\Hil$ is cyclic and seperating for each algebra $\A(W)$. Moreover,
the modular operators of these algebras with respect to $\Omega$ are
known to act geometrically correct, {\em i.e.} as expected from the
Bisognano-Wichmann theorem \cite{BuLe}.\\
In the above construction, we distinguished $W_L$ as reference wedge
by generating the algebras associated to left wedges $W_L+x$ by the
fields $\phi(\psi)$, $\psi\in\LL(W_L+x)$, and defined the algebras
associated to right wedges as the corresponding commutants. Another possible
definition of the wedge algebras, which distinguishes $W_R$ instead of
$W_L$ as reference wedge, is given by interchanging $W_L$ and $W_R$ in
\bref{aw}. Note that these two definitions do {\em not\/}
coincide. With the convention \bref{aw} used here, $\A(W_R+x)$ is not generated by
$\phi(\psi)$, but by a second, different field $\phi'(\psi)=S\phi(\psi)S^*$,
lying relatively wedge-local to $\phi$ \cite{gl1}, {\em i.e.}
\begin{eqnarray}
  \A(W_R+x)=\{S\phi(\psi)S^*\,:\,\psi\in\LL(W_R+x)\}''\,.
\end{eqnarray}
But since $S$ \bref{S} commutes with the translations, the two nets arising
from distinguishing either $W_L$ or $W_R$ as reference wedge are
unitarily equivalent.  Therefore we can adopt the convention \bref{aw}
without loss of generality.\\
To analyze the content of local observables of the net \bref{aw} of wedge
algebras, we consider a double cone $\OO_x := W_L \cap (W_R+x)$,
$x\in W_L$ and the algebra
\begin{equation}\label{ao}
  \A(\OO_x) := \A(W_L) \cap \A(W_R+x)\,.
\end{equation}
According to the modular nuclearity condition, the existence of
observables localized in $\OO_x$ is ensured if the map
\begin{eqnarray}\label{xil}
  \Xi(x) : \A(W_L) \lto \Hil,\qquad \Xi(x)(A) :=
  \Delta^{1/4}U(x)A\Omega\,,
\end{eqnarray}
is nuclear, where $\Delta$ denotes the modular operator of
$(\A(W_L),\Omega)$. Comparing with section two, we
see that $\Xi(x)$ has the form of the map $\Xi_\LL$ considered there
with $\LL=\LL(W_L+x)$ and $X =
\Delta^{1/4}$. Indeed, $\Delta^{1/4}$ is closed, strictly positive,
and is the seond quantization of its restriction to the one particle
space. One-particle states $\psi\in\dom\Delta^{1/4}$ have wavefunctions admitting an analytic continuation to the strip
$S(0,\frac{\pi}{2}) = \{\te\in\Cl\,:\,0<\mbox{Im}(\te)< \frac{\pi}{2}\}$
with continuous boundary values, and one
has in particular 
\begin{equation}
  (\Delta^{1/4}\psi)(\te) = \psi(\te+\tfrac{i\pi}{2})\,.
\end{equation}
Consequently, $\Delta^{1/4}$ commmutes with $\Gamma$. 
We therefore may apply
Proposition \ref{prop} to deduce the nuclearity of the map \bref{xil}
from the nuclearity of the one-particle operators 
\begin{equation}\label{tc}
  T_\varphi(x) := \Delta^{1/4}E_\varphi(W_L+x),\qquad\;\; T_\pi(x) :=
  \Delta^{1/4}E_\pi(W_L+x), \qquad x\in W_L, 
\end{equation}
where $E_\varphi(W_L+x), E_\pi(W_L+x) \in \B(\K)$ denote the
orthogonal projections on the subspaces $\LL_\varphi(W_L+x)$ and
$\LL_\pi(W_L+x)$, respectively.
But the operators \bref{tc} were already shown to be of trace
class on $L^2(\Rl,d\te)$ in \cite{BuLe}. We have thus verified the
modular nuclearity condition in this model, thereby finishing the
construction of the corresponding net of local
algebras \bref{ao}.\\
To summarize, the map $\OO\longmapsto\A(\OO)$
defined by \bref{ao} and \bref{aw} is an isotonous, local net of von
Neumann algebras which is covariant with respect to the action of the
representation $U$ because of the corresponding properties of the net
of wedge algebras \cite{gl1}. Furthermore, we established here the following Proposition. 
\begin{proposition}
  In the model theory with S-matrix $S=(-1)^{N(N-1)/2}$, the maps
  $\Xi(x)$ are nuclear for any $x\in W_L$. As a consequence, the local
  algebras $\A(\OO)$ \bref{ao} are isomorphic to the hyperfinite type
  III$_1$ factor for any double cone $\OO$.\\
  In particular, every local algebra $\A(\OO)$ has cyclic vectors and
  therefore contains non-trivial operators.{\hfill $\square$ \\[2mm] \indent}
\end{proposition}

\section{Conclusions}
\setcounter{equation}{0}

We have verified the modular nuclearity condition in a factorizing
model with non-trivial S-matrix, thus realizing the algebraic
construction of such theories in a first example. The work presented
in this article is another step in the constructive
program initiated by Schroer \cite{schroer,schroer-wiesbrock} which
provides a further test of the utility of the modular nuclearity
condition in this context. Applying the results from
\cite{BuLe}, the local algebras in this model were shown to be
non-trivial and satisfy all postulates of nets of local
observables. One may therefore expect that this algebraic approach is
also viable in models with a more general factorizing S-matrix. The
nuclearity properties of such theories will be discussed elsewhere.\\
In comparison to other treatments of factorizing models \cite{BKW,
  mccoy,ffp}, it is apparent that the approach described here and the
formfactor program are complementary to each other: Whereas
explicit constructions of $n$-point functions and computations of
formfactors are more convenient in the field-theoretic
context, the algebraic approach appears to be better suited for
the discussion of existence problems. It seems that one needs both approaches for a proper understanding of this
area of quantum field theory.\\
In conclusion we mention that the methods developed here can also be
applied to theories of free Fermions in order to establish nuclearity
properties. Consider a free Fermi field on $(d+1)$-dimensional
Minkowski space, and a double cone $\OO_r := \{(x_0,x)\in\Rl^{1+d}\,:\,|x_0|+|x| < r\}$. Let $\A(\OO_r)$ denote the corresponding von Neumann algebra of
observables localized in $\OO_r$. Fixing a Hamiltonian $H$ as the
generator of translations along some timelike direction, we consider
the maps 
\begin{eqnarray}
  \Theta_{\beta,r} : \A(\OO_r)\lto\Hil,\qquad \Theta_{\beta,r}(A) := e^{-\beta H}A\Omega\,,\qquad\quad\beta>0.
\end{eqnarray}
Analogous to $\Xi(x)$ \bref{xil}, $\Theta_{\beta,r}$ can be formulated in
terms of a real subspace of the one particle space \cite{BuWi}. Setting
$X=e^{-\beta H}$, the mathematical structure described in section two
is thus seen to be present also in this context. Hence Proposition
\ref{prop} can be applied to reduce the nuclearity problem for
$\Theta_{\beta,r}$ to a question on the one particle space, which is
known to have an affirmative answer \cite{BuWi}. Applying the
calculations from the appendix of \cite{BuWi}, one can also derive
bounds on the nuclear norms $\|\Theta_{\beta,r}\|_1$ in terms of $\beta$, the
spacetime dimension and the diameter $r$ of the localization region considered, thereby 
establishing the energy nuclearity condition and all of
its consequences \cite{BuWi,BuJu} in this class of models.

\subsection*{Acknowledgements}
Many thanks are due to D. Buchholz for numerous helpful discussions and
constant support. I would also like to thank K.-H. Rehren for useful suggestions improving
the manuscript. This work has been funded by the Deutsche Forschungsgemeinschaft (DFG).

\end{document}